\documentclass[
 aip,
 jmp,
preprint,
reprint,
author-year,
author-numerical,
]{revtex4-1}
\usepackage{times} 
\usepackage{graphicx}
\usepackage{dcolumn}
\usepackage{bm}
\usepackage{changes}
\usepackage{textcomp}
\usepackage[utf8]{inputenc}
\usepackage[T1]{fontenc}
\usepackage{mathptmx}
\usepackage{etoolbox}
\usepackage{ragged2e}
\usepackage{booktabs}
\usepackage{physics}
\usepackage{tabularx} 
\usepackage{xcolor}
\usepackage{titlesec}
\usepackage[]{units}
\usepackage[]{upgreek}

\DeclareUnicodeCharacter{0308}{\"}
\DeclareUnicodeCharacter{030C}{\v{ }}
\DeclareUnicodeCharacter{0301}{\'{ }}

\newcommand{\sinc}{\text{sinc}}
\usepackage{mathtools} 

\UseRawInputEncoding 

\makeatletter
\def\@email#1#2{%
 \endgroup
 \patchcmd{\titleblock@produce}
  {\frontmatter@RRAPformat}
  {\frontmatter@RRAPformat{\produce@RRAP{*#1\href{mailto:#2}{#2}}}\frontmatter@RRAPformat}
  {}{}
}%
\makeatother

\usepackage{times} 

\titleformat{\section}[block]
{\normalfont\rmfamily\selectfont\bfseries\large}
  {\thesection}{1em}{}
\titlespacing*{\section}{0pt}{\baselineskip}{\baselineskip} 

\titleformat{\subsection}[block]
{\normalfont\rmfamily\selectfont\bfseries\normalsize}
  {\thesubsection}{1em}{}
\titlespacing*{\subsection}{0pt}{0pt}{\baselineskip} 

\makeatletter
\renewcommand{\fnum@figure}{\textbf{Fig. \thefigure}}
\makeatother
\begin{document}

\preprint{AIP/123-QED}

\title{\rmfamily\bfseries\LARGE Nonreciprocal spin waves in out-of-plane magnetized waveguides reconfigured by domain wall displacements}

\author{Hanadi Mortada} \affiliation{Fachbereich Physik and Landesforschungszentrum OPTIMAS, Rheinland-Pf\"alzische Technische Universit\"at Kaiserslautern-Landau, 67663 Kaiserslautern, Germany}

\author{Roman Verba} \affiliation{Institute of Magnetism, Kyiv 03680,
Ukraine.}

\author{Qi Wang} \affiliation{School of Physics, Huazhong University, Wuhan, China.}

\author{Philipp Pirro} \affiliation{Fachbereich Physik and Landesforschungszentrum OPTIMAS, Rheinland-Pf\"alzische Technische Universit\"at Kaiserslautern-Landau, 67663 Kaiserslautern, Germany}

\author{Alexandre Abbass Hamadeh*}
\affiliation{Universit\'e Paris-Saclay, Centre de Nanosciences et de Nanotechnologies, CNRS, 91120, Palaiseau, France}

\email{alexandre.hamadeh@universite-paris-saclay.fr}

\date{\today}

\begin{abstract}
 Wave-based  platforms for novel unconventional computing approaches like neuromorphic computing require a well-defined, but adjustable flow of wave information combined with non-volatile data storage elements to implement  weights which allow for training and learning. Due to their inherent nonreciprocal properties and their direct physical interaction with magnetic data storage, spin waves are ideal candidates to realize such platforms. In the present study, we show  how spin-wave nonreciprocity induced by dipolar interactions of nanowaveguides with antiparallel, out-of-plane  magnetization orientations can be used to create a spin-wave circulator allowing for unidirectional information transport and complex signal routing. In addition, the device can be reconfigured by a magnetic domain wall with adjustable position, which allows for a non-volatile tuning of the nonreciprocity and signal propagation. These properties are demonstrated for a spin-wave directional coupler through a combination of micromagnetic simulations and analytical modeling also showing that it functions as a waveguide crossing element,  tunable power splitter, isolator, and frequency multiplexer. As magnetic material, out-of-plane magnetized Bismuth-doped Yttrium Iron Garnet has been considered. For this material, the motion of domain walls by magnonic spin transfer torque has been recently experimentally demonstrated which enables to store results from spin-wave computation. In combination with the presented concept of domain wall based reconfiguration and nonlinear spin-wave dynamics, this enables for the creation of a nano-scaled nonlinear wave computing platform with the capability for self-learning. 
\vspace{0.2cm}
\end{abstract}
\maketitle

\noindent Magnonic networks, which represent a set of interconnected micro- and nanoscale magnetic structures, are expected to play a pivotal role in advancing the next-generation of functional devices for un-conventional computing concepts, microwave and terahertz information and communication systems, all grounded in magnonic principles \cite{chumak2022advances,wang2024nanoscale,Pirro2021Advancescoherent}. The evolution of such networks from passive wave-based conduits into intelligent, reconfigurable components offers a transformative opportunity for neuromorphic hardware systems. The dipolar interactions in nanoscale spin-wave (SW) couplers not only facilitate efficient information routing but also open the door to embedding computational logic directly within the physical dynamics of wave propagation\cite{szulc2025magnetic,sadovnikov2017voltage,sadovnikov2015directional,wang2021inverse,wang2018reconfigurable,sadovnikov2017toward}. Previous proposals for SW device connectors have explored the use of adjacent dipolarly coupled waveguides/stripes \cite{sadovnikov2017voltage,odintsov2024nonreciprocal,odintsov2024lateral} and nano-sized directional couplers\cite{szulc2020spin,szulc2025magnetic,sadovnikov2017spin,sadovnikov2015directional,odincov2020intensity,sadovnikov2018spin,wang2019nanoscale,wang2020nonlinear,wang2020magnonic}, emphasizing the impact of waveguide structure and SW mode interactions on coupling efficiency. 
Experimental and numerical studies have delved into the nonreciprocal propagation characteristic of SWs utilized in magnonic technologies \cite{grachev2024nonreciprocal,khalili2007nonreciprocal,odintsov2024nonreciprocal,flebus20242024,mruczkiewicz2013nonreciprocity,mruczkiewicz2014observation,gladii2016frequency,di2013spin,grassi2020slow}. Nonreciprocity-defined as the directional asymmetry in SW transmission-introduces a powerful mechanism to modulate wave-based signal flow. Such asymmetric transport is not only essential for traditional microwave devices like isolators and circulators, but is increasingly being recognized as a fundamental building block in magnonic logic and interconnect networks\cite{millo2023unidirectionality,szulc2020spin,grunberg2008nobel,jamali2013spin,chen2021unidirectional,wang2021nonreciprocal,deorani2014nonreciprocity,zenbaa2025yig,grassi2020slow} making magnonic systems uniquely promising for unconventional computing platforms, such as neuromorphic and reservoir computing architectures \cite{menezes2024toward,monalisha2024magnetoionics}. In such systems, computing emerges from the spatiotemporal dynamics of wave interference, propagation delays, and adaptive weights-requirements that are naturally addressed by SWs due to their phase coherence, frequency tunability, and rich non-linear interactions. Critically, nonreciprocal SW propagation can be exploited to implement asymmetric signal routing, emulate synaptic asymmetry, and enable spatial segregation of excitation and readout-a key requirement for energy-efficient, hardware-level learning. Moreover, SWs can interact with magnetic textures like domain walls (DWs), which serve as non-volatile, reconfigurable elements encoding computational states. This interaction allows for the realization of self-adaptive systems in which SWs both process information and dynamically reconfigure the medium, embodying the core principle of self-learning: co-localized computation and memory. 

While predominantly low-damping, in-plane magnetized waveguides have been extensively used to interconnect magnonic channels \cite{grachev2023reconfigurable,wang2018reconfigurable,zhao2022reconfigurable,odintsov2024nonreciprocal}, there has been limited exploration of the potential in normally magnetized materials to achieve controlled directivity and manipulation of dipolarly coupled SWs\cite{zhang2022spin,zhao2022reconfigurable,ren2019reconfigurable,hamalainen2018control}. This gap presents a compelling opportunity to explore nonreciprocal SW transport in systems exhibiting strong perpendicular magnetic anisotropy (PMA), where the magnetic texture can be further reconfigured using DWs that themselves are manipulable via propagating SWs\cite{fan2023coherent}. A particularly promising candidate is bismuth-doped yttrium iron garnet (Bi:YIG), a low-damping insulating ferrimagnet that enables out-of-plane magnetization in films up to 140 nm thick-significantly surpassing the thickness constraints of undoped YIG due to the strain-induced limitations of achieving PMA\cite{das2023perpendicular,raja2021enhanced,dastjerdi2021magnetic,soumah2018ultra,kumar2019ferromagnetic,fakhrul2019magneto,qinghui2006study,butler1990microwave,das2023perpendicular}. Recent experimental advances have shown that coherently excited SWs in Bi:YIG can act as spin current carriers to drive and manipulate DWs with high energy efficiency and long propagation distances \cite{fan2023coherent}, establishing Bi:YIG as a highly favorable material for integrated magnonic logic systems. These systems inherently decouple charge from information transport, instead combining SWs for signal processing with magnetic DWs for non-volatile memory, enabling a magnetic analog to logic-in-memory computing \cite{pirro2023coherent}.Consequently, the material parameters of Bi:YIG will be considered in our study to construct the proposed devices further detailed. Nevertheless, other doped YIG variants like Ga:YIG could be used to realize similar concepts. 

To render a directional coupler based on dipolar interactions between laterally adjacent SW waveguides viable for modern magnonic circuits, the device must meet several stringent criteria: nanoscale compactness, operation without external bias fields, low propagation losses (due to both intrinsic damping and interfacial reflections), and seamless integration into larger circuits without relying on additional transduction layers. In this work, we address these challenges by investigating forward volume spin waves (FVSWs) in normally magnetized, self-biased, laterally parallel Bi:YIG nanowaveguides that are dipolarly coupled. Utilizing a combination of micromagnetic simulations and analytical theory, we demonstrate that our directional coupler-comprising two bent nanoscale waveguides closely positioned to maximize magnetostatic interaction-exhibits robust nonreciprocal behavior and multifunctionality. Governed by tunable parameters such as frequency, waveguide geometry, separation, and magnetization orientation, the coupler performs critical tasks including power division, frequency multiplexing, signal isolation, and waveguide crossing. Its pronounced nonreciprocity emerges from gyrotropic magnetization precession and the time-reversal symmetry breaking in the antiparallel out-of-plane magnetization configuration. Crucially, we introduce a novel reconfiguration mechanism based on the insertion of a magnetic DW into one of the waveguides. The DW position directly modulates the nonreciprocal transmission properties, offering a non-volatile and spatially tunable degree of control that is absent in previous implementations. This capability enables the construction of reconfigurable SW devices in which the flow of energy can be dynamically guided by the DW position. Such DW-tunable control serves as a basis for non-volatile memory elements essential for advanced magnonic data processing platforms, including neuromorphic computing. Furthermore, in Bi:YIG, DWs can be displaced by intense SW bursts through magnonic spin-transfer torque (STT), as demonstrated experimentally \cite{fan2023coherent}. This opens the door to device concepts where SW information is not only transmitted but also transduced into a non-volatile state variable-the DW position-which can later be accessed by additional SWs. This self-reconfigurable behavior, wherein the SWs themselves influence the functional state of the device, constitutes a foundational prerequisite for enabling unsupervised or self-learning processes in nanoscale, wave-based computational systems.

\begin{figure*}[t]
  \centering
  \includegraphics[scale = 1.1]{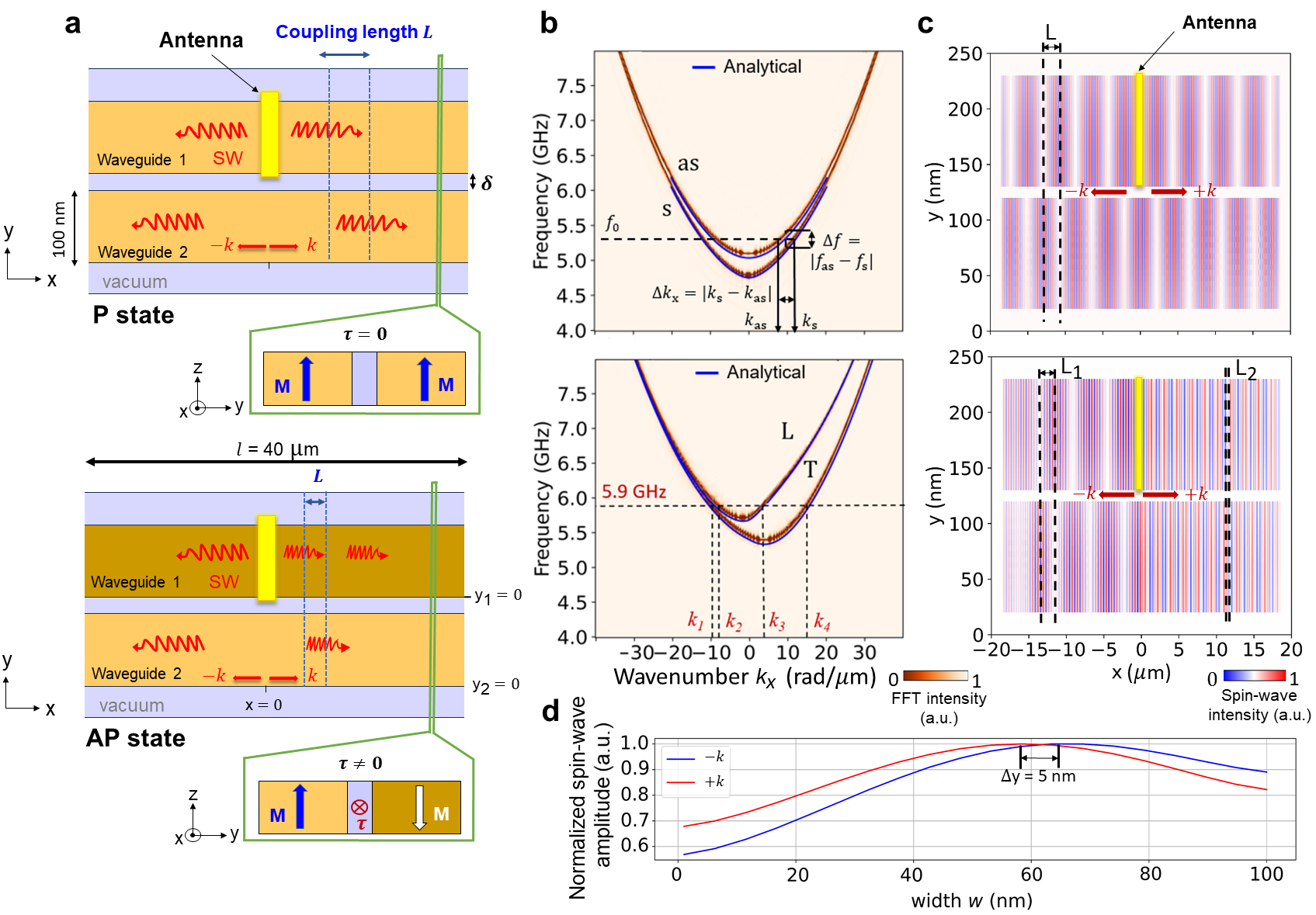}
  \caption{\textbf{Nonreciprocity of FVSWs and dependence on the static magnetization configuration.}  \textbf{(a)} An illustrative depiction of top view and a cross section of dipolarly coupled SW waveguides with parallel (P) and antiparallel (AP) static magnetizations. The excitation antenna (shown in yellow) is positioned at the center of waveguide 1. A toroidal moment $\tau$ along $\textbf{k}$ is only present in the AP case, suggesting the unique occurrence of nonreciprocity along the x-direction. \textbf{(b)} The corresponding dispersion curves for the coupled waveguides in the P and AP states. The results of numerical simulations are depicted using a color map, while the blue lines correspond to the predictions of the analytical theory. The dispersion curves in the P configuration is split into antisymmetric (as) and symmetric (s) branches whereas in the AP configuration, nonreciprocal longitudinal (L) and transverse (T) branches emerge. \textbf{(c)} A top view snapshot displaying the SW profile stimulated within the P and AP coupled waveguides at the excitation frequency of 5.9 GHz. \textbf{(d)} SW profile along the width of waveguide 1 in the AP configuration for positive and negative wavenumbers extracted from a line scan at x = $-5$ $\mu$m and x = $+5$ $\mu$m. A weak localization of the waves near one edge of the waveguide, depending on their propagation direction, is observed.}   
  \label{fig:1}
\end{figure*}

\section*{Results}
\subsection*{Spin-wave nonreciprocity for out-of-plane magnetized waveguides in antiparallel configuration }

\noindent Previous research has demonstrated that when two identical magnetic SW waveguides are placed in close proximity, their dipolar interaction induces a splitting of modes in the waveguides resulting in the formation of a symmetric "s" and an antisymmetric "as" mode. For the "s" mode, the magnetic oscillations in the two waveguides are in phase (at the same x-position, see Fig. 1), and the "as"  mode is characterized by a 180$^{\circ}$ phase difference\cite{wang2018reconfigurable}. If these collective modes are excited simultaneously at a certain frequency, the dipolar coupling leads to periodic energy transfer between the waveguides. Consequently, the energy initially excited in one waveguide is transferred to the other after traversing a specific distance, termed the coupling length $L = \frac{\pi}{\Delta k_x} = \frac{\pi}{\lvert k_s - k_{as} \rvert}$ where \(k_s\) and \(k_{as}\) are the wavenumbers for the symmetric and antisymmetric modes at the excitation frequency, respectively\cite{sasaki1981directional}. For a given length of the coupled waveguides, the coupling length determines how energy is distributed between the different outputs. To analytically characterize the energy transfer in the system of two coupled waveguides, it is imperative to first determine the dispersion relation and the spatial profiles of the SW eigenmodes within them. The explicit derivation of the characteristics of the collective modes which necessitates solving the Landau-Lifshitz-Gilbert (LLG) equation of magnetization dynamics can be found in the Supplementary Methods 1. In a bias-free system, waveguides can exhibit two stable static magnetic configurations parallel "P" and antiparallel "AP" static magnetizations, which profoundly affect the coupled modes and the coupling length.  In the "AP" case, symmetry-breaking phenomena arise from the gyrotropic motion of the magnetization vector as that of both waveguides rotate in opposite directions and the SWs propagate perpendicular to the static magnetization. This leads to a unique spatial distribution of the dipolar field generated by their dynamic magnetizations and thus, the characterization of the internal field in the dispersion relation for two coupled waveguides with AP magnetization configuration exhibits nonreciprocity. This is reflected in the dispersion relation :

\begin{equation}
\begin{split}
\omega_{1,2}(k_x) = \sqrt{(\Omega^{xx} \pm \omega_M F_{k_x}^{xx}(d_{12}))(\Omega^{yy} \mp \omega_M F_{k_x}^{yy}(d_{12}))}\\ + \text{sign}(k_x) \omega_M \Im{F_{k_x}^{xy}(d_{12})}
\end{split}
\end{equation}

\noindent where we assume the static magnetizations $\mathbf{\mu_1} = + \textbf{e}_z$, $\mathbf{\mu_2} = - \textbf{e}_z$, and $d_{12}$ is defined as $d_{12} = y_{c,1}-y_{c,2}$, such that $y_{c,i}$ is the coordinate of the waveguide's center. The spectrum exhibits nonreciprocity, as indicated by the final term in equation (1). The function $\text{sign}(k_x)$ ensures that the last term explicitly depends on the propagation direction of the SW. This effect is significant and scales linearly with the coupling strength. The dynamic magneto-dipolar interaction between the coupled waveguides, characterized by the off-diagonal element of the dipole-dipole interaction, results in a stronger interaction between the oppositely precessing dynamic magnetizations\cite{wang2018reconfigurable,szulc2020spin} breaking the SW reciprocity with opposite signs of wavevectors ($\pm k$), i.e. the amplitude and the modal profile of counter-propagating SWs differ considerably. In this configuration, "s" and "as" modes are absent; instead, the collective modes are characterized by asymmetric $x$ and $y$-polarized (or longitudinal "L" and transverse "T") modes. 

In this study, the two materially and geometrically  identical parallel normally-magnetized waveguides, each with a length of 40 $\mu$m, a width $w$ of 100 nm and a thickness $h$ of 50 nm, are placed adjacent to each other with a gap $\delta$ of 10 nm (see Fig. 1(a)). The numerical modeling of these structures is carried out using the MuMax3 micromagnetic software package\cite{vansteenkiste2014design} and the Bi:YIG film material parameters are assigned as: uniaxial anisotropy
constant $K_u$ = 12.5 $\text{ kJ/m}^3$ , exchange stiffness $A_{\text{ex}} = 4.2  \, \text{ pJ/m}$, Gilbert damping $\alpha = 1.3 \times 10^{-4}$, and saturation magnetization $M_s$ = 100 $\text{kA/m}$\cite{fan2023coherent}.  No external bias magnetic field is applied. To initiate the propagation of the SW in one of the waveguides, a time-varying sinusoidal magnetic field \( b_y = b_0 \sin(2\pi ft) \) is applied in a 10 nm wide region around the center ($x$ = 0) with an oscillation amplitude $b_0$ = 1 mT and a varying microwave frequency $f$. Due to the low Gilbert damping of Bi:YIG, SWs can propagate over significant distances, thereby enabling the realization of intricate integrated SW circuits. Fig. 1(b) illustrates the dispersion relations of the two traveling modes for the aforementioned pair of coupled waveguides with parallel and opposite static magnetization $M_{z1}, M_{z2}$ along the z-direction. The dispersion relations corresponding to our numerically simulated coupled waveguides reveal a splitting of the fundamental width mode within each isolated waveguide prior to coupling, resulting in the emergence of two distinct modes due to dipolar interactions when the waveguides are brought into close proximity. This phenomenon has been extensively investigated in previous studies involving in-plane magnetized YIG \cite{wang2018reconfigurable, wang2019nanoscale, wang2020magnonic,wang2024nanoscale}, where propagating SW modes in adjacent waveguides exhibit reciprocal behavior, satisfying \( f(k) = f(-k) \). However, a pronounced nonreciprocal effect is observed in the AP configuration, where \( f(k) \neq f(-k) \), with the wave vectors \( k_{1,2} \) and \( k_{3,4} \) differing by nearly a factor of two at \( f = 5.9 \) GHz. This nonreciprocity is  characteristic for out-of-plane magnetized systems and is absent in the previously studied in plane systems \cite{wang2018reconfigurable} .

Spin-wave nonreciprocity is conveniently characterized through the definition of a toroidal moment $\tau$ \cite{gorbatsevich1994toroidal,zimmermann2014ferroic}, contingent on the equilibrium magnetization $\textbf{M}$ and wave vector $\textbf{k}$. In particular, nonreciprocal SW propagation is realized when the condition $\tau \cdot \textbf{k} \neq 0$ is satisfied, implying that nonreciprocity emerges when a toroidal moment component is oriented parallel to the propagation direction ($\tau \parallel k$) \cite{foggetti2019magnetic}. Consequently, the toroidal moment provides a predictive framework for identifying dipolar symmetry breaking ($f(k)\ne f(-k)$) within the magnon dispersion relation. As illustrated in the waveguide's cross-sections in Fig. 1(a), in the AP state, there is no symmetry operation connecting a reciprocal situation since $\tau \ne 0$ and \textbf{$\tau || k$} , while in the P state, $\tau = 0$. The minor discrepancies between the numerical results and the analytical theory arise from internal field inhomogeneities, which are not accounted for in the analytical model (the local effective field maxima at the inner edges are diminished in comparison to those at the outer edges, resulting in a shift of the modes profiles towards the inner edges and an increase in coupling strength).
\vspace{0.2cm}

Snapshots capturing the dynamic magnetization profiles within the coupled waveguides excited at 5.9 GHz are shown in Fig. 1(c) illustrating the clear difference between the periods of energy exchange between the waveguides along the positive and negative propagating directions in the P and AP state. The SW powers in both waveguides, in scenarios where realistic SW damping is considered, gradually diminishes and can be expressed as $P_{1} = P_{in} \sin^2 \left( \frac{\pi x}{2L} \right) \exp \left( -\frac{|2x|}{x_{\text{free path}}} \right)$  where $P_1$ is the power in waveguide 1 in which the antenna is positioned, $x_{\text{free path}}$ is the exponential decay length, and $P_{in}$ is the input SW power in the waveguide 1. A single coupling length $L$ = 2.22 $\mu$m can be determined in the P configuration along opposite propagation directions. However, in the AP case, the two directions show very different coupling,  with $L_1$ = 2.22 $\mu$m along the $-k$ direction and a much stronger coupling  along the $+k$ direction ($L_2$ = 0.27 $\mu$m). The difference of the coupling lengths is due to the different wavenumbers in opposite propagation directions  (2.98 rad/$\mu$m and 14.6 rad/$\mu$m for the longitudinal mode and transverse mode in the $+k$ direction, respectively, and $-8.79$ rad/$\mu$m for the longitudinal mode and $-9.58$ rad/$\mu$m for the transverse mode in the $-k$ direction). The modes' group velocities slightly differ as well along $\pm k$ where, for the longitudinal mode, a group velocity of $v_{g-L}$ = 35.5 nm/ns and $-$44.9 nm/ns is recorded in the $+k$ and $-k$ direction, respectively. Similarly, for the transverse mode, $v_{g-T}$ = 77.8 nm/ns and $-$74.6 nm/ns are recorded. The presence of off-diagonal dynamic components in the demagnetization tensor constitutes the primary distinction of FVSW behavior in coupled waveguides with out-of-plane magnetization when compared to in-plane magnetized SW coupled waveguides. Fig. 1(d) displays the result of the numerical simulation of the SW profiles across the width of waveguide 1 in the AP state extracted at x = $-5$ $\mu$m and x = $+5$ $\mu$m. A small shift of approximately $\Delta y$ = 5 nm is distinctly apparent in the profiles, suggesting that the FVSW within the nanoscale waveguide exhibits a nonreciprocity characteristic similar to the Damon-Eshbach surface waves \cite{wang2021inverse}.\\

\subsection*{Domain wall implementation}

\begin{figure*}[t]
  \centering
  \includegraphics[scale = 1.1]{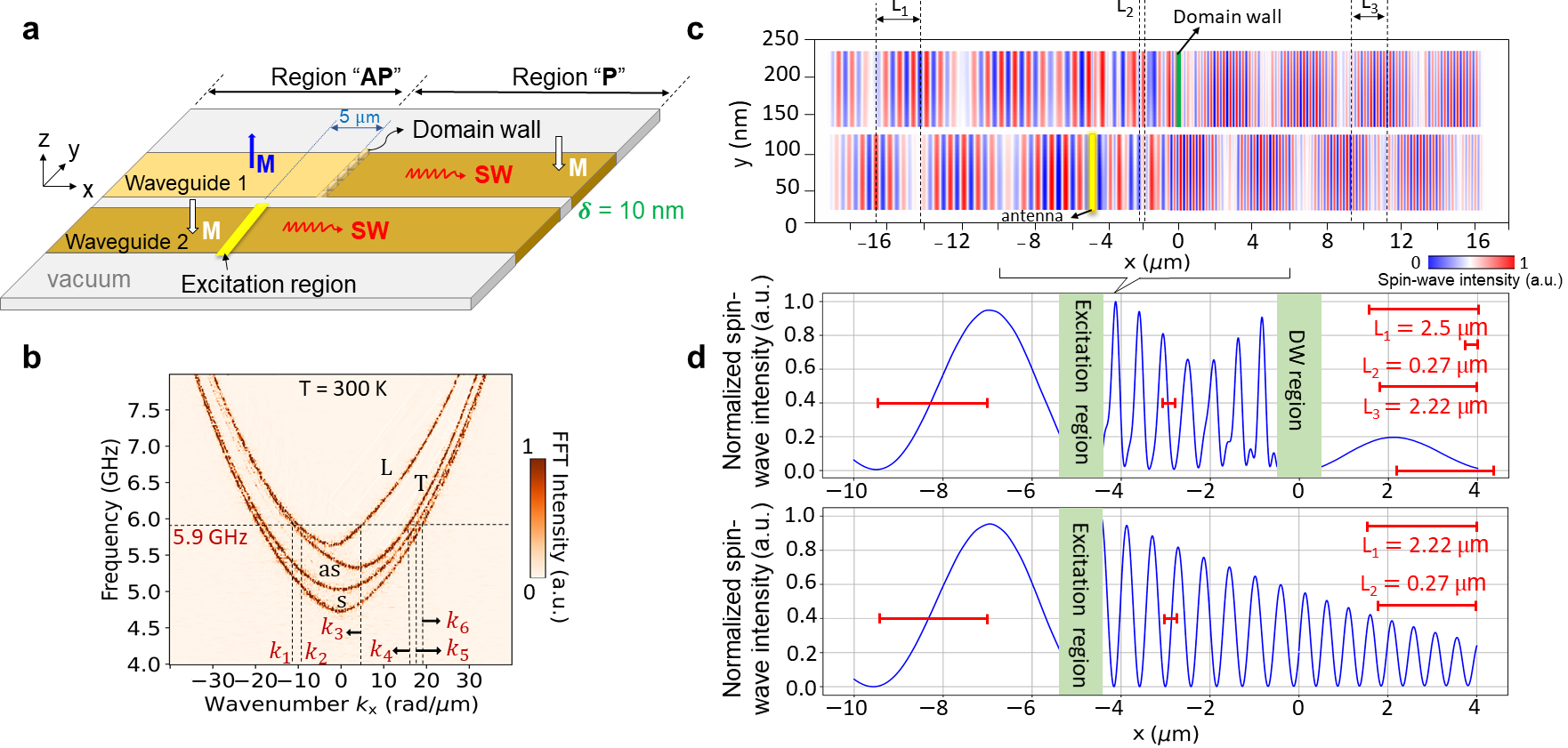}
  \caption{\textbf{Domain wall implementation in dipolarly coupled waveguides.} \textbf{(a)} An illustrative depiction of dipolarly coupled SW waveguides including a DW. The excitation antenna (shown in yellow) is positioned 5 $\mu$m away from the center of the second waveguide.  \textbf{(b)} Dispersion curves for the split modes propagating in the coupled waveguides of parallel and antiparallel regions with SW excited by thermal agitation. \textbf{(c)} A top view snapshot displaying the SW profile stimulated with DW implemented at x = 0 at $f$ = 5.9 GHz.   \textbf{(d)} Spin-wave propagation recorded in waveguide 1 along the area encompassing the regions surrounding the DW, both in the presence and absence of a DW.}   
  \label{fig:2}
\end{figure*}

\noindent YIG thin films typically do not exhibit PMA due to YIG's weak magnetocrystalline and magnetoelastic anisotropies, which fail to surmount the waveguide's in-plane shape anisotropy. Thus, re-magnetizing the waveguides independently via external magnetic fields necessitates the application of a brief magnetic field pulse perpendicular to the SW propagation direction, followed by its reduction to zero, thereby engendering a N\'eel DW at the center of the structure\cite{wang2018reconfigurable}. Conversely, in materials with pronounced PMA, such as Bi:YIG, the magnetization naturally aligns out-of-plane, facilitating the formation of narrow Bloch walls. These systems are thus more adept at controlling and exploiting DWs for nanoscale magnonic applications. If a waveguide, bearing a DW at its midpoint ($x_{DW}$ = 0), is proximally positioned to another waveguide, the resultant configuration is dissected into two distinct regions, one of a P magnetization configuration and another manifests an AP state, as illustrated in Fig. 2(a). Consequently, the averaged SW dispersion relation in this arrangement encompasses modes similar to those observed in DW-free coupled waveguides with P and AP states, culminating in four modes at a single frequency. The dispersion relation within these coupled waveguides, excited at ambient room temperature (T = 300 K) (see to Fig. 2(b)), delineates the nonreciprocal nature of the AP region along the $\pm k$ directions and the symmetry of the SW dispersion in the P region propagating in the $+k$ direction. At $f$ = 5.9 GHz, the nonreciprocal wavenumbers in the AP region of the waveguides, exhibit values of $k_1$ and $k_4$ for the transverse modes, and $k_2$ and $k_3$ for the longitudinal modes in the $−k$ and $+k$ propagation directions, respectively. Conversely, $k_5$ and $k_6$ denote the wavenumbers for the antisymmetric and symmetric modes, respectively, in the P region. A snapshot of the dynamic magnetization profiles within the coupled waveguides excited by an alternating magnetic field at 5.9 GHz (simulated at at T = 0 K), depicts the variation in the period of energy exchange as the SW traverses the DW, revealing interference between incident and reflected SWs in the pre-DW region (see to Fig. 2(c)). The antenna is placed in waveguide 2, 5 $\mu$m away from the DW which is positioned at the center of waveguide 1. As the SW is excited in the AP region, the energy of the initially excited SW in waveguide 2 is completely transferred to waveguide 1 over the coupling length in a nonreciprocal manner, with $L_1$ = $\frac{\pi}{\lvert k_1 - k_2 \rvert}$ = 2.5 $\mu$m along the $-k$ direction and $L_2$ =$\frac{\pi}{\lvert k_3 - k_4 \rvert}$ = 0.27 $\mu$m in the $+k$ direction. A comparative analysis of the SW intensity in the vicinity of the antenna, both with and without the DW, reveals a 11.42$\%$ reflection of the incident wave energy at the DW boundary for $f$ = 5.9 GHz for the coupled waveguides' system. The oscillatory nature of the reflected wave is observed to be dependent on the specific spatial arrangement of the DW.  For an isolated waveguide with a DW implemented, the spin-wave transmits through the DW with a 4.5$\%$ reflection rate at $f$ = 5.9 GHz. However, this reflection coefficient can be substantially reduced by increasing the excitation frequency and decreasing the ratio of the SW wavelength to the DW width. For instance, utilizing a frequency of 7.5 GHz yields a reflection coefficient of only 2.95$\%$ for a dipolarly-coupled waveguide and of 0.4$\%$ for an isolated waveguide. This behavior can be attributed to the resonant characteristics of the system, where certain frequencies interact more effectively with the DW, resulting in a reduced reflection at the boundary. The attenuated transmission of the SW in the P region of the waveguides is exchanged with a periodicity of $L_3$ =$\frac{\pi}{\lvert k_5 - k_6 \rvert}$ = 2.22 $\mu$m (see Fig. 2(d)).
\vspace{0.2cm}

\subsection*{Fundamental operating concept and multi-functionality of the directional coupler}
\noindent Dipolarly coupled SW modes within parallel waveguides present considerable potential for various applications. The proposed directional coupler in this paper comprises two nanosized, self-biased SW Bi:YIG waveguides with bent regions positioned laterally close to each other, thereby establishing a region for dipolar coupling as illustrated in Fig. 3. The specific shape of the bent segments ensures highly efficient SW transmission. In this simulation, the waveguides' width was maintained at $w$ = 100 nm, thickness at $h$ = 50 nm, and the gap between waveguides at $\delta$ = 10 nm. The length of the coupled waveguides is set to $l$ = 6 $\mu$m.
The functionality of the device based on two laterally parallel coupled waveguides depends on the ratio between the coupling length ($L$) and the total length of the coupled waveguides ($l$). In the absence of a DW, the ratio of the output power in the upper waveguide to the aggregate power, denoted as $\frac{P_{2}}{P_{2} + P_{4}}$, can be expressed in terms of the characteristic coupling length, disregarding damping effects, as follows:

\begin{equation}
  \frac{P_{2}}{P_{2} + P_{4}} = \cos^2 \left( \frac{\pi l}{2L} \right)
\end{equation}

\noindent Here, \( P_{2} = P_{1} \cos^2 \left( \frac{\pi l}{2L} \right) \) and \( P_{4} = P_{1} \sin^2 \left( \frac{\pi l}{2L} \right) \) represent the output powers for waveguides 1 and 2, respectively, and \( P_{1} \) is the input SW power in the first waveguide.
\vspace{0.2cm}

\begin{figure}[t]
  \centering
  \includegraphics[scale = 0.8]{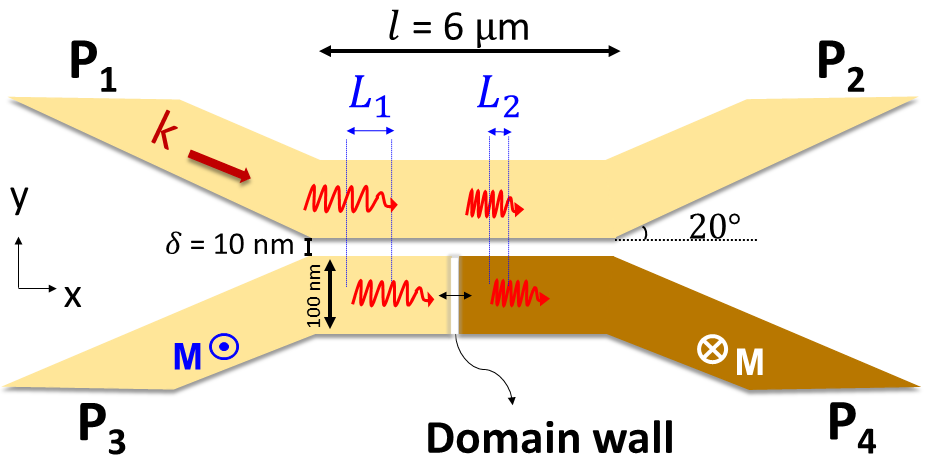}
  \caption{\textbf{Versatile SW Directional Coupler.} A schematic view of the proposed four-port directional coupler with the implementation of a DW at the center waveguide 2 to be displaced. The widths of the waveguides are $w$ = 100 nm, thickness is $h$ = 50 nm, and gap is $\delta$ = 10 nm; the angle between the coupler waveguides is 20$^{\circ}$; and the working length of the coupled waveguides is $l$ = 6 $\mu$m. The SWs are excited in the first beam pass of the directional coupler marked as "$P_{1}$".}   
  \label{fig:3}
\end{figure}

\begin{figure*}[t]
  \centering
  \includegraphics[scale = 0.8]{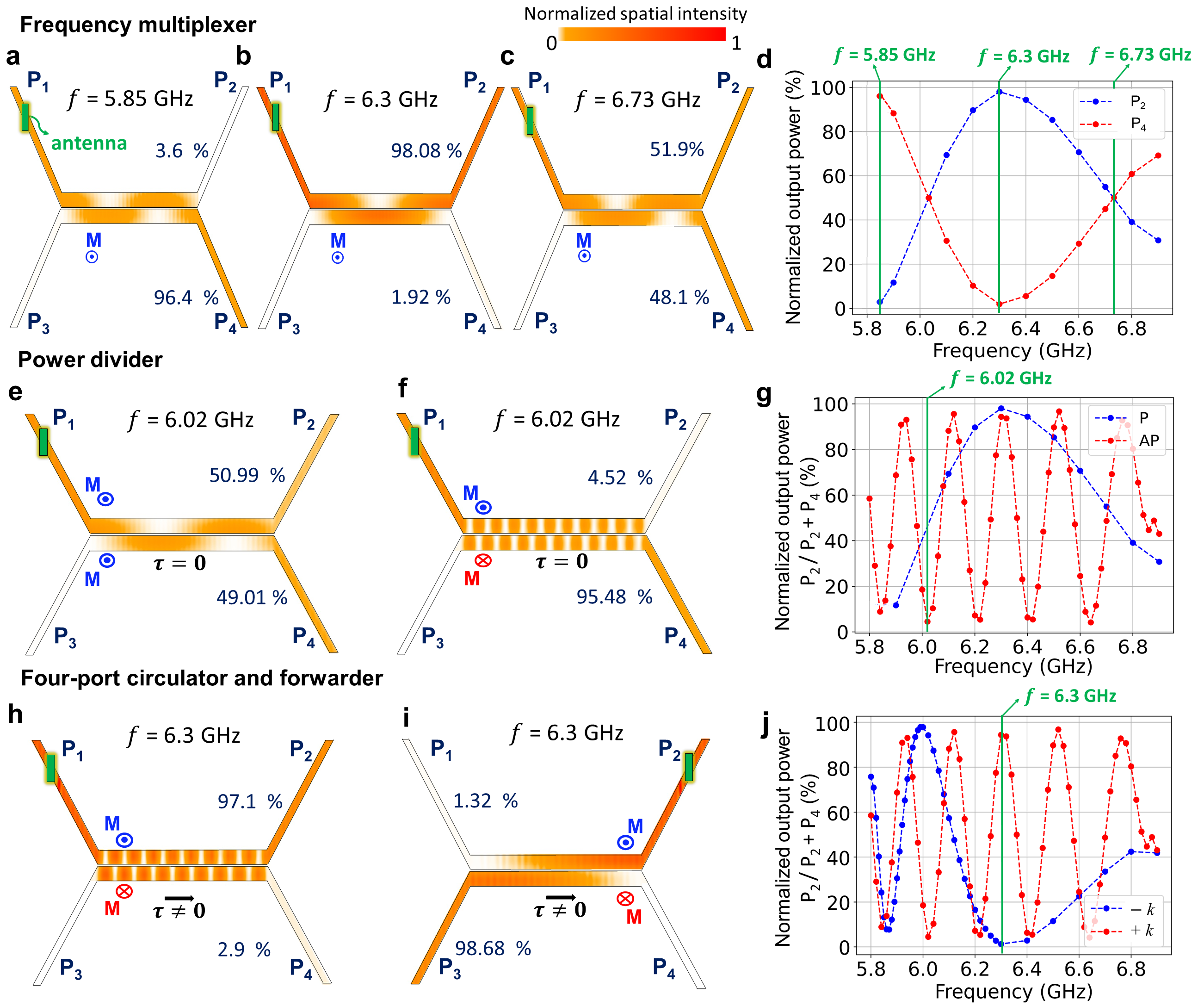}
  \caption{\textbf{Adjustable SW directional coupler.} \textbf{(a to c)} Switching of the device functionality by changing the signal frequency. A directional coupler acts as a connector of magnonic conduits, as a power divider, or as a simple transmission line (delay line). The SW intensity is shown by a color map. Note that the width of the waveguides is constant in all parts of the directional coupler, as shown in Fig. 3. \textbf{(d)} Frequency dependence of the normalized power at the output of the first and second beam pass in the directional coupler. \textbf{(e and f)} Switching of the device's functionality by changing the static magnetization orientation. Switching the relative orientation of the static magnetization in two beam passes leads to the switching of the output signal between the beam passes. \textbf{(g)} Frequency dependence of the normalized power at the output of the parallel (P) and antiparallel (AP) configuration beam passes in the directional coupler. \textbf{(h and i)} Switching of the device functionality by changing the excitation direction. \textbf{(j)} Frequency dependence of the normalized power at the output of the antiparallel (AP) configuration beam passes in the directional coupler with opposite excitation directions, $+k$ and $-k$.}   
  \label{fig:4}
\end{figure*}

\noindent If $l=(2n + 1)L$ where $n$ is an integer, the entirety of the energy transfers from one waveguide to the other, rendering the directional coupler apt for linking magnonic conduits. Conversely, if $l=(n + 1/2)L$, the coupler can function as an equal divider. As established in prior research on YIG, the operational characteristics of the directional coupler can be regulated by modifying the geometric dimensions of the waveguides, adjusting their interspacing, varying the SW frequency, and altering the relative orientations of static magnetizations in the interacting SW waveguides \cite{wang2018reconfigurable}.  However, the use of Bi:YIG introduces an additional reconfiguration mechanism-DW displacement within one of the waveguides. Together with the possibility to use magnonic spin-transfer torque to achieve the necessary DW displacement, this possibility to reconfigure the magnonic device allows for interesting concepts like unsupervised learning \cite{menezes2024toward,wang2024nanoscale}. The angle between the sections of the directional coupler is fixed at 20$^{\circ}$, sufficiently small to ensure efficient SW transmission through the bend in the SW waveguide. The operational characteristics of the directional coupler varies with the input microwave signal frequency. Also, it can be reconfigured by adjusting: the static magnetization configuration (P and AP state), the excitation site (i.e., the selected input port utilizing either $+k$ or $-k$ direction for SW propagation), and the DW position, thereby providing an additional degree of flexibility in employing the Bi:YIG coupler.
\\

\noindent \textit{Changing the functionality based on frequency and magnetization orientation} 
\vspace{0.2cm}

\noindent This section serves to validate the operational mechanism and design of the proposed device by examining its reconfigurability through variations in excitation frequency and static magnetization. A comprehensive analysis of this functionality, particularly in the context of in-plane magnetized YIG, can be found in the prior work of Wang. et al\cite{wang2018reconfigurable}. Figs. 4(a), (b), and (c) present color maps illustrating the normalized spatial intensity distribution of SWs within the directional coupler. Fig. 4(d) shows the normalized output power in the first and second beam paths, represented by $\frac{P_{2}}{P_{2} + P_{4}}$ and $\frac{P_{4}}{P_{2} + P_{4}}$, for the P state configuration of the waveguides' magnetization. For excitation frequencies of \( f = 5.85 \) GHz and \( f = 6.3 \) GHz, the majority of the SW energy is directed toward a single output port ($P_4$ and $P_2$, respectively), as the propagation length \( l \) corresponds to an integer multiple of the coupling length (\( l = 3L \) at 5.85 GHz and \( l = 2L \) at 6.3 GHz), thereby enabling the coupler to operate as a frequency-selective separator or multiplexer. Consequently, when SWs of two distinct frequencies (5.85 GHz and 6.3 GHz) are simultaneously excited, the coupler effectively differentiates between them, directing each signal to its designated output port. Furthermore, Fig. 4(c) demonstrates the dual functionality of the device as a power divider, wherein energy is symmetrically distributed between \( P_{2} \) and \( P_{4} \). Specifically, for an SW propagating at \( f = 6.73 \) GHz, the coupling condition \( l = 2.5L \) is satisfied. As previously discussed, altering the static magnetization configuration from P alignment to an AP alignment also affects the coupling between the waveguides. For instance, when a SW is excited at a frequency of 6.02 GHz (see Fig. 4(e) and (f)), a substantial portion of the SW energy is transferred from $P_{1}$ to $P_{2}$ in the AP state due to a significantly reduced coupling length, satisfying the criterion $l = 20 \times L$, in contrast to the P state, as demonstrated in Fig. 4(g). This finding highlights the dynamic adaptability of the proposed directional coupler, which can effectively function as a rapid switch and/or multiplexer. The operational frequency of the device can be fine-tuned by adjusting the length of the coupled region $l$, by modifying the waveguide geometry to alter the coupling length $L$ or by applying an external bias magnetic field.

\begin{figure*}[t]
  \centering
  \includegraphics[scale = 1.2]{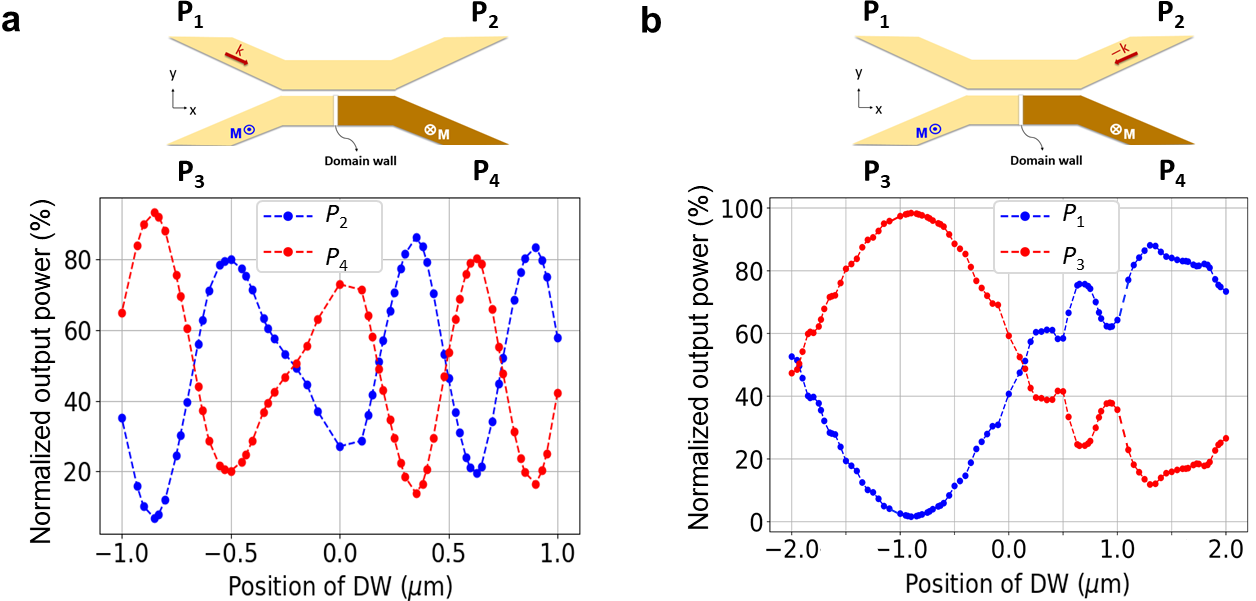}
  \caption{\textbf{Domain Wall modulating parameter of function.}  A schematic view of the directional coupler with the implementation of a domain wall to be displaced and the corresponding DW position dependence of the normalized power at the output ports \textbf{(a)}  \( P_{2} \) and  \( P_{4} \) excited along the $+k$ direction and \textbf{(b)} \( P_{1} \) and  \( P_{3} \) excited along the $-k$ excitation direction in the directional coupler at 5.9 GHz.}  
  \label{fig:5}
\end{figure*}
\vspace{0.2cm}

\noindent \textit{Changing the functionality based on the SW nonreciprocity in the antiparallel configuration} 
\vspace{0.2cm}

\noindent The dipolar interaction between waveguides in an AP configuration disrupts the symmetry of SW dispersion. By leveraging the nonreciprocal nature of SWs, it is possible to develop SW isolators and circulatory elements. These functionalities can be directly demonstrated by altering the propagation direction of the incident SW from \( +k \) to \( -k \) (see Fig. 4(j)). At an excitation frequency of \( f = 6.3 \) GHz, when the SW is excited along the positive propagation direction (\( +k \)), the majority of the wave power is transmitted to the first output port (\( P_2 \)) (Fig. 4(h)). Conversely, when the SW is injected in the negative propagation direction (\( -k \)), most of the wave power is directed to the second output port (\( P_4 \)) (Fig. 4(i)). This behavior stems from the difference in the \( l/L \) ratio in the lateral configurations at \( f = 6.3 \) GHz, where \( l = 18 \times L \) in Fig. 4(h) and \( l = 2 \times L \) in Fig. 4(i). This additional degree of control, which is not attainable in previously proposed single-layer systems, introduces a novel functionality in Bi:YIG. It enables the device to operate as a port-forwarder, where energy injected at \( P_1 \) propagates sequentially to \( P_2 \), but energy inject at  \( P_2 \) is directed to \( P_3 \). This feature significantly enhances the device's versatility and applicability in advanced magnonic and microwave circuits. Furthermore, the device is capable of functioning as a four-port circulator, where energy introduced at \( P_1 \) transitions to \( P_2 \) (\( +k \) excitation), but from \( P_2 \) is guided to \( P_3 \) (\( -k \) excitation),  from \( P_3 \) to \( P_4 \) (\( +k \) excitation), and finally from \( P_4 \) back to \( P_1 \) (\( -k \) excitation).
The ability to precisely manipulate power and information flow direction of SWs is a prerequisite for the creation of many data processing schemes, including, e.g., forward feeding (neural) networks, where direction-dependent signal flow is essential for implementing asymmetric weight functions and mimicking synaptic transmission. In such systems, nonreciprocity ensures that information travels preferentially along predefined pathways, allowing the network to encode causality and suppress signal reflection or feedback that could disrupt learning dynamics.

\vspace{0.2cm}

\noindent \textit{Changing the functionality based on domain wall displacements}
\vspace{0.2cm}

\noindent Introducing a DW into one of the waveguides within the directional coupler creates regions with parallel or oppositely directed static magnetizations (see Fig. 5). As a result, the system shows a kind of "effective coupling length" $L_{eff}$ which is a function of the coupling length in the P and AP state as well as of the position of the DW (compare Fig. 2).  Consequently, the position of the DW allows to tune the  $l/L_{eff}$ ratio. As a result, the DW position alters the ratio of output power between waveguide 1 and waveguide 2. At a specific frequency $f$ = 5.9 GHz, the coupling lengths in the excitation region are measured as $L_{parallel}$ = 1.99 $\mu$m and 2.5 $\mu$m for incident propagation directions $+k$ and $-k$, respectively. These values change to $L_{anti-parallel}$ = 0.27 $\mu$m and 2.22 $\mu$m in these directions after passing through the DW. Fig. 5 illustrates the intricate dependence of the normalized power at the different outputs of the device at a fixed frequency of 5.9 GHz, with DW displacements ranging from a minimum of $x_{DW}$ = $-2 \mu$m to a maximum of $x_{DW}$ = 2 $\mu$m from the center of the coupled waveguides. Due to the nonreciprocity of the system, the power distribution depends also on the propagation direction ($+k$ in 5(a) and $-k$ in 5(b)) similar to the results presented in Fig. 4(h). The reconfigurable guiding of the SW energy as a function of the DW position allows to construct non-volatile memory elements for advanced magnonic data processing schemes like neuromorphic computing. 
Such control mirrors synaptic plasticity in biological systems, where signal transmission strength is modified in response to previous stimuli. More profoundly, the DW position itself can be altered by intense SW bursts \cite{fan2023coherent}, allowing the SWs not only to carry information but also to modify the functional state of the device where information can be transferred to a the non-volatile variable (DW position) which can be read out later by other SW.. This dynamic interplay-wherein the wave both programs and reads its own environment-paves the way for inherently self-reconfigurable elements that embody key attributes of unsupervised learning. By embedding memory and computation into the same physical substrate, such devices offer a foundational building block for scalable, energy-efficient magnonic neuromorphic architectures. 

\section*{Discussion}

\noindent A versatile directional coupler was proposed, and its functionality as a waveguide
crossing element, tunable power splitter, controlled multiplexer, and frequency separator was demonstrated. This apparatus operates via the interference of two collective SW modes within laterally aligned and dipolarly coupled magnetic waveguides, with a gap between them. The investigation delved into the coupling length, denoted as $L$, representing the distance over which SW energy transfers between coupled waveguides. This coupling length underwent scrutiny concerning SW frequency, the geometry of the coupler, and the relative alignment of static magnetization in waveguides. Our micromagnetic simulations have unveiled two critical advantages of the proposed device. First, it exploits the intrinsic nonreciprocity stemming from the antiparallel static magnetization of the waveguides, which enables highly directional SW propagation. This nonreciprocal behavior is fundamental to the device's functionality, as it ensures robust signal routing and minimizes undesired backscattering, a key requirement for efficient wave-based information processing. Second, the device offers dynamic reconfigurability through the controlled displacement of DWs within the interconnected structure. By shifting the DWs-potentially via SWs-the coupling properties can be actively modified, allowing the device to switch between different operational states without the need for external magnetic fields. This ability to reconfigure its functionality on demand adds a significant degree of versatility, making the directional coupler highly adaptable for nanoscale microwave signal processing in both digital and analog applications. Our micromagnetic simulations verify the feasibility of actualizing the device experimentally. While the results are promising, additional research is necessary to improve the comprehension of spin wave-domain wall interactions in these devices. Although this research paper primarily employed theory and simulations to design the device and demonstrate its functionality and operational mechanism, experimental studies are ongoing to build a proof of concept. Moreover, this material has demonstrated effectiveness in the manipulation of magnetic DWs due to its minimal magnon attenuation, using coherently excited SWs as spin current carriers. These discoveries contribute to a deeper comprehension of the interplay among spin waves generated by diverse inputs in a directional coupler setup. Our micromagnetic simulations verify the feasibility of actualizing the device experimentally. 

\section*{Methods}
\subsection*{An analytical theory of coupled waveguides}
\noindent 
Deriving the characteristics of the collective modes necessitates solving the Landau-Lifshitz (LL) equation of magnetization dynamics
\begin{equation}
\frac{d\mathbf{M}}{dt} = -|\gamma| \mathbf{M} \times \mathbf{B}_{\text{eff}}
\end{equation}

\noindent in the linear approximation while neglecting the damping term. Here, \(\mathbf{M}\) denotes the magnetization vector, \(\mathbf{B}_{\text{eff}}\) represents the effective field, and \(\gamma\) is the gyromagnetic ratio. For the purposes of this analysis, uniform thickness profile is assumed, which is a valid approximation for waveguides with thicknesses on the order of a hundred nanometers or less. Examining two identical SW waveguides and FVSW modes propagating along these waveguides in the \( x \) direction, the magnetization in a waveguide can be expressed as $ \mathbf{M}(\mathbf{r}, t) = M_s(\bm{\mu}$ $ + \mathbf{m}(y)\exp(i(kx + \omega_k t)))$, where \( M_s\) is the saturation magnetization, $\bm{\mu}$ is the unit vector in the direction of the static magnetization, and \( \mathbf{m} \ll 1 \) represents the small dynamic deviation of the magnetization from its equilibrium position. Utilizing this form in equation (3), the SW frequency $\omega_k$ and vector structure $\mathbf{m_\textit{\textbf{k}}}_{,p}$ are determined as
\begin{equation}
-i\omega_{k_x} \mathbf{m}_{k_x,p}= \bm{\mu}_p \times \sum_q \hat{\Omega}_{k_x, pq} \mathbf{m}_{{k_x},q}
\end{equation}

\noindent which is the Fourier representation (in time and \( x \) coordinate) of the linearized LL equation. Here, $\bm{\mu}_p = \pm \textbf{e}_z$ is the unit vector along the direction of the static magnetization; indices $p, q = 1, 2$ enumerate the waveguides; and the tensor $\hat{\Omega}_{k_x, pq}$ is given by:
\begin{equation}
\hat{\Omega}_{k_x, pq} = \gamma \left( B_p + \mu_0 M_s \lambda_{ex}^2 k_x^2 \right) \delta_{pq} \hat{\textbf{I}} + \omega_M \hat{\textbf{F}}_{k_x} (d_{pq})
\end{equation}

\noindent where $B_p$ is the static internal magnetic field (encompassing exchange fields, uniaxial anisotropy field, static demagnetizing fields, and external fields, if any), \( \delta_{pq} \) is the Kronecker delta function, \( \delta_{pq} = 0 \) if \( p \ne q \), \( \delta_{pp} = 1 \), \( d_{pq} \) is the distance between the centers of the two waveguides (\( d_{pp} = 0 \), \( d_{12} = -d_{21} = w + \delta \)), \( \delta \) is the gap between the waveguides of width \( w \), \( \lambda_{ex} = \sqrt{2A/(\mu_0 M_s^2)} \) is the exchange length, \( \omega_M = \gamma \mu_0 M_s \), \( \mu_0 \) is the vacuum permeability, and \( \hat{\textbf{I}} \) is the identity matrix. Advised that the general dispersion relation and spatial configuration of SWs for waveguides with out-of-plane and in-plane magnetization are identical. The impact of the dynamic magneto-dipolar interaction, both within and between waveguides, as described in equation (5), is encapsulated by the tensor \(\hat{\textbf{F}}_{k_x}\) developed as:
\begin{equation}
\hat{\textbf{F}}_{k_x}(d_{pq}) = \frac{1}{2\pi} \int \hat{\textbf{N}}_\textbf{\textit{k}} e^{ik_y d_{pq}} \, d k_y
\end{equation}
with
\begin{equation}
\hat{\textbf{N}}_\textbf{\textit{k}} = \frac{|\sigma_{k}|^2}{w}
\begin{pmatrix}
\frac{k_x^2}{k^2} f(kh) & \frac{k_x k_y}{k^2} f(kh) & 0 \\
\frac{k_x k_y}{k^2} f(kh) & \frac{k_y^2}{k^2} f(kh) & 0 \\
0 & 0 & 1 - f(kh)
\end{pmatrix}
\end{equation}

\noindent where $f(kh) = 1 - \frac{1 - \exp(-kh)}{kh}\), \(\textbf{\textit{k}} = k_x \textbf{\textit{e}}_x + k_y \textbf{\textit{e}}_y $ $(k = \sqrt{{k_x}^2 + {k_y}^2}$, \(h\) denotes the waveguide thickness, \(\sigma_k = w \sinc(k_y w/2)\) represents the Fourier transform of the SW profile across the waveguide width that is assumed to be uniform (\( m(y) = 1 \)), which occurs if the waveguide size is comparable to or smaller than the material's exchange length \( \lambda_{ex} \) or if the effective boundary conditions are free. The magnitude of the static internal magnetic field is given by 
\begin{equation}
B_p = \mathbf{B}_e .\bm{\mu}_p + B_a - \mu_0 M_s (F_0^{zz}(0) + F_0^{zz}(d_{pq})\bm{\mu}_p. \bm{\mu}_q)
\end{equation}
 
\noindent where $\mathbf{B}_e$ is out-of-plane external field (if applied) and $B_a$ is the uniaxial anisotropy field, which is supposed to be along the z-direction.
When \(d_{pq}=d_{pp} = 0\), the tensor \(\hat{\textbf{F}}_{k_x}(0)\) corresponds to the dipolar self-interaction within the waveguide, while \(\hat{\textbf{F}}_{k_x}(d_{pq})\) represents the dipolar interaction between waveguides. $\hat{\textbf{F}}_{k_x}$ possesses the following properties: $\hat{\textbf{F}}_{k_x}(0)$ is real (as long as the static magnetization is directed along one of the symmetry axes of the waveguide) and diagonal, $\hat{\mathbf{F}}_{k_x}(d_{pq}) = \hat{\mathbf{F}}^*_{k_x}(-d_{pq}) = \hat{\mathbf{F}}^*_{-k_x}(d_{pq})$, with real diagonal components and imaginary equal off-diagonal xy and yx components. The presence of off-diagonal dynamic components in the demagnetization tensor constitutes the primary distinction of FVSW behaviour in coupled waveguides with out-of-plane magentization when compared to an in-plane magnetized SW coupled nanowires. Owing to the effective dipolar boundary conditions at the lateral boundaries of the waveguides, the width profiles of the collective SW modes tend to exhibit partial pinning \cite{guslienko2002effective, guslienko2005boundary}. This phenomenon results in a non-uniform width profile of the fundamental SW mode of the waveguide, denoted as \( m(y) \sim \cos(\kappa y) = \cos\left(\frac{\pi y}{w_{\text{eff}}}\right) \), where \( w_{\text{eff}} \) represents the effective width of the waveguide. This non-uniformity is characterized by the effective wave number \( \kappa \) such that the effective width \( w_{\text{eff}} = \frac{\pi }{\kappa} \) can be substantially larger than the nominal waveguide width \( w \) when the degree of effective pinning is reduced. The normalization constant of the mode profile \( m(y) \), denoted as \( \tilde{w} \), is defined by $\tilde{w} = \int_{-w/2}^{w/2} m(y)^2 \, dy$. In instances where the SW profile is nearly uniform (i.e., $w_{\text{eff}} \rightarrow \infty)$, $ \tilde{w} = w$. For other spatially symmetric modes with the profile \( m(y) \sim \cos(\kappa y) \), the calculation of dynamic components of the magneto-dipolar tensor in equation (7) after interchanging $w$ by $ 
\tilde{w} = \frac{w}{2} \left( 1 + \text{sinc}(\kappa w) \right)
$ results in
\begin{equation}
\begin{split}
\sigma_k = \frac{2}{{k_y}^2-{\kappa}^2} &\left[ k_y \cos\left( \frac{\kappa w}{2} \right) \sin\left( \frac{k_y w}{2} \right) \right.\\
&\left. - \kappa \cos\left( \frac{k_y w}{2} \right) \sin\left( \frac{\kappa w}{2} \right) \right]
\end{split}
\end{equation}

\noindent Additionally, in equation (5), an extra exchange term emerges, $\lambda_{ex}^2k_{x}^2 \rightarrow \lambda_{ex}^2(k_{x}^2 + \kappa^2)$. Also, a modification pertains to the definition of the static demagnetization tensor, which is utilized for the static field calculation in equation (8), necessitating the substitution of $|\sigma_{\textbf{\textit{k}}}|^2 /w$ with $(\sigma_k^{(0)})^* \sigma_k^{(2)} / \tilde{w}$ where
\begin{equation}
\sigma_k^{(n)} = \int_{-w/2}^{w/2}\cos^n(\kappa y) e^{ik_y y} dy
\end{equation}

\noindent The dispersion relation in an isolated nanowaveguide is
\begin{equation}
\omega_0(k_x) = \sqrt{\Omega^{xx}\Omega^{yy}}
\end{equation}
where
\begin{equation}
\Omega^{\alpha\alpha} = \omega_H + \omega_M(\lambda^2(k_x^2 + \kappa^2)+ F_{k_x}^{\alpha\alpha}(0))
\end{equation}

\noindent The bias-free version of equation (12) denotes
\begin{equation}
\omega_H = \gamma B_a-\omega_M F_0^{zz}(0)
\end{equation}

\noindent The dispersion relation for two coupled waveguides with parallel magnetization configuration "\textbf{P}", accounting for the symmetry properties of the tensor $\hat{\textbf{F}}_{k_x}$, can be derived as 
\begin{equation}
\begin{split}
\omega^2_{1,2}(k_x) = &\ \Omega^{xx}\Omega^{yy} + \omega^2_M(F_{k_x}^{xx}F_{k_x}^{yy} - {|F_{k_x}^{xy}|}^2) \\
& \hspace{-2cm} \pm \sqrt{\omega^2_M(F_{k_x}^{xx}\Omega^{yy} + F_{k_x}^{yy}\Omega^{xx})^2 - 4\omega^4_MF_{k_x}^{xx}F_{k_x}^{yy}{|F_{k_x}^{xy}|}^2}
\end{split}
\end{equation}

\noindent illustrating the dispersion reciprocity where, for simplicity, we denote $\hat{\textbf{F}}_{k_x} \equiv \hat{\textbf{F}}_{k_x}(d_{12})$. When the magneto-dipolar interaction between waveguides is negligible in comparison to the dipolar self-interaction within a single waveguide, the dispersion relations for the two collective modes can be approximated by 
\begin{equation}
{\omega}_{1,2}(k_x) \approx {\omega}_0(k_x) \pm \omega_M \frac{\Omega^{xx}F_{k_x}^{yy}(d_{12})+ \Omega^{yy}F_{k_x}^{xx}(d_{12})}{2\omega_0(k_x)}
\end{equation}

\noindent which closely resembles that of an in-plane coupler \cite{wang2018reconfigurable}. It is noteworthy that the $F_{k_x}^{xy}$ component does not significantly influence the coupled mode dispersion in the initial approximation. Additionally, in the calculations given by equations (14) and (15), one must incorporate the mutual static demagnetization, 
\begin{equation}
{\omega}_{H}=\gamma B_a - \omega_M(F_0^{zz}(0)+F_0^{zz}(d_{12}))
\end{equation}

\noindent when calculating $\Omega^{\alpha\alpha}$ and $\omega_0(k_x)$. In the bias-free scenario, a simple dispersion relation for two coupled waveguides with antiparallel  magnetization configuration "\textbf{AP}" can be derived as
\begin{equation}
\begin{split}
\omega_{1,2}(k_x) = \sqrt{(\Omega^{xx} \pm \omega_M F_{k_x}^{xx}(d_{12}))(\Omega^{yy} \mp \omega_M F_{k_x}^{yy}(d_{12}))}\\ + \text{sign}(k_x) \omega_M \Im{F_{k_x}^{xy}(d_{12})}
\end{split}
\end{equation}

\noindent where we assume the static magnetizations $\mathbf{\mu_1} = + \textbf{e}_z$, $\mathbf{\mu_2} = - \textbf{e}_z$, and $d_{12}$ is defined as $d_{12} = y_{c,1}-y_{c,2}$, such that $y_{c,i}$ is the coordinate of the waveguide's center. In the approximation of weak coupling, a simplified equation is obtained
\begin{equation}
\begin{split}
{\omega}_{1,2}(k_x) \approx \omega_0(k_x) \pm \omega_M \frac{\Omega^{yy}F_{k_x}^{xx}(d_{12})- \Omega^{xx}F_{k_x}^{yy}(d_{12})}{2\omega_0(k_x)} \\\pm \omega_M\Im{F_{k_x}^{xy}(d_{12})}
\end{split}
\end{equation}

\begin{figure}[b]
  \centering
  \includegraphics[scale = 1.2]{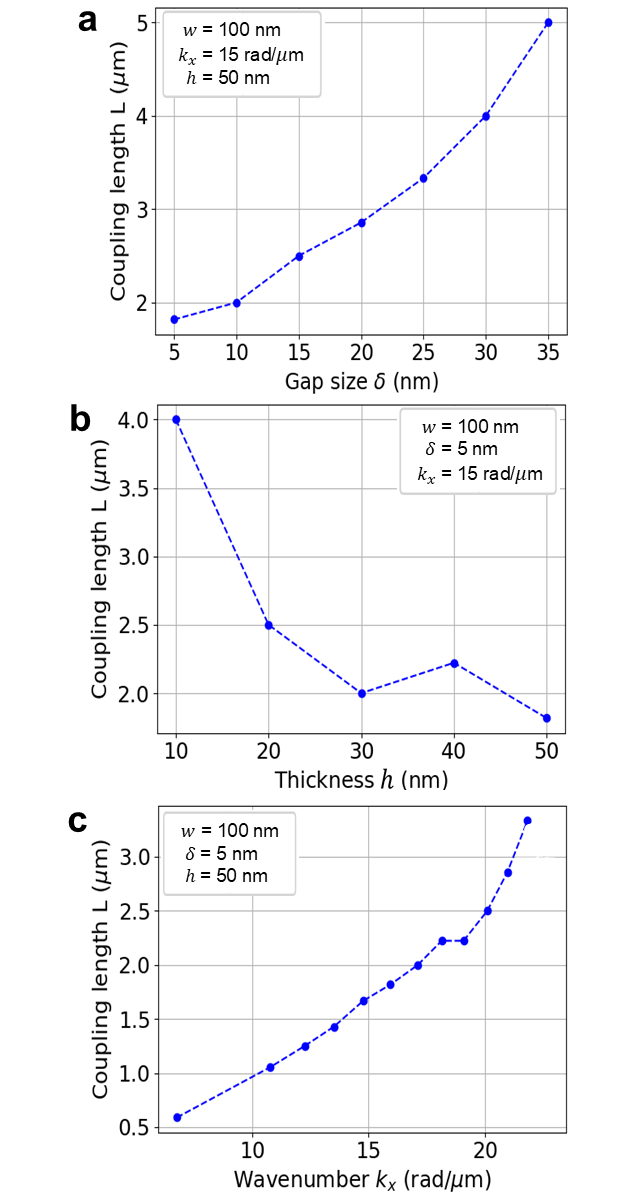}
  \caption{\textbf{Coupling length.} The dependence of the coupling length $L$ illustrated in terms of \textbf{(a)} the gap separation $\delta$ between the waveguides. \textbf{(b)} The thickness $h$ of the waveguides. \textbf{(c)} The longitudinal SW wavenumber $k_x$. The specific parameters for each case are provided within the respective panels.}   
  \label{fig:6}
\end{figure}

\noindent The spectrum exhibits nonreciprocity, as indicated by the final term in equation (17). This nonreciprocity can be effectively described by introducing the toroidal moment $\tau$-a vectorial multipole quantity that captures the closed-loop, head-to-tail circulation of magnetic moments, inherently breaking both time-reversal and spatial-inversion symmetries, and acting as a fundamental order parameter in systems exhibiting nonreciprocal behavior-defined in relation to the equilibrium magnetization $\textbf{M}$ and the spin-wave wave vector $\textbf{k}$. Both equations (14) and (17) describe the relationship between the frequency separation of the collective modes and the various geometrical and material parameters of a system comprising two coupled waveguides. Generally, the frequency separation, denoted as \( \Delta f \), increases with the enhancement of dynamic magneto-dipolar interactions between the waveguides. This enhancement can be achieved by reducing the inter-waveguide separation or by increasing the thickness-to-width ratio of the waveguides. Additionally, \( \Delta f \) is highly sensitive to the saturation magnetization of the materials employed. The developed analytical theory provides a relatively straightforward and intuitive method for calculating the characteristics of magnonic spin wave couplers, in contrast to conventional micromagnetic simulations. Note that in our study, the excited SW is not a classical FV magnetostatic wave (FVMSW) like in a plane film since the exchange energy is also taken into account in this case. Comparing the numerical results to the 
analytical theory, based on the approximation of uniform dynamic magnetization, shows an overestimation of the coupling due to the complex static demagnetization field landscape (the local effective field maxima at the inner edges are diminished in comparison to those at the outer edges, resulting in a shift of the modes profiles towards the inner edges and an increase in coupling strength).  Consequently, unlike in the BVSW case \cite{wang2018reconfigurable}, any semi-analytical theory applied to FVSW geometry should be regarded as an approximation, necessitating the use of numerical or micromagnetic methods for precise calculations.\\

\subsection*{Coupling Length}

The coupling length $L$ depends on various external and internal system parameters such as the type of the material, the geometrical sizes of the waveguides (i.e. the gap size $\delta$, the thickness $h$), their static magnetization, the SW wavenumber, and the bias magnetic field (if applied). 
Typically, an increase in dipolar interaction between waveguides results in a reduction of $L$. Fig. 6 shows the coupling length dependence on these parameters for laterally parallel waveguides. It is noticed that the coupling is intensified by reducing the gap, $\delta$, between the waveguides (Fig. 6(a)) or increasing their thickness, $h$ (Fig. 6(b)). Additionally, it is observed that the coupling strength diminishes as SWs of higher wavenumbers, $k_x$, are excited  due to reduced dipolar interaction for short-wavelength SWs (see to Fig. 6(c)). The coupling length determines how energy is distributed between the outputs over a specific length of the coupled waveguides, thereby influencing the output power. Note that the Gilbert damping is set to exponentially increase to 0.8 at the ends of the waveguides to prevent backscattering and standing waves' formation.

\section*{References}
\vspace{-3em}
\bibliography{references}
\bibliographystyle{naturemag}

\section*{Author contributions}
\vspace{-1em}
\noindent H.M. performed micromagnetic simulation, and wrote the first version of the manuscript. A.H. and P.P. devised and planned the project. R.V. developed the general analytical theoretical model. All authors discussed the results and contributed to writing the manuscript. 
\section*{Competing interests}
\vspace{-1em}
\noindent The authors have no conflicts to disclose.
\section*{Additional information}
\vspace{-1em}
\noindent \textbf{Correspondence} and requests for materials should be addressed to Alexandre A. Hamadeh.
\end{document}